\documentclass{article}

\usepackage[final, nonatbib]{neurips_2022_ml4ps}

\usepackage[utf8]{inputenc} % allow utf-8 input
\usepackage[T1]{fontenc}    % use 8-bit T1 fonts
\usepackage{hyperref}       % hyperlinks
\usepackage{url}            % simple URL typesetting
\usepackage{booktabs}       % professional-quality tables
\usepackage{amsfonts}       % blackboard math symbols
\usepackage{nicefrac}       % compact symbols for 1/2, etc.
\usepackage{microtype}      % microtypography
\usepackage{xcolor}         % colors
\usepackage{graphicx}      % handle figures
\usepackage{subfig}         % handle sub-figures
\usepackage{xfrac}          % typeset fractions cleanly
\usepackage[backend=biber,sorting=none,style=ieee,citestyle=numeric-comp]{biblatex}
\iffalse
\newcommand{\nsection}[1]{\section{#1}}

\else

\newcommand{\ncaption}[1]{\vspace{-0.15cm}\caption{#1}{\vspace{-0.4cm}}}
\newcommand{\nsection}[1]{\vspace{-0.1cm}\section{#1}\vspace{-0.1cm}}

\fi

\title{Geometry-aware Autoregressive Models for Calorimeter Shower Simulations}

\author{
  \parbox[c]{0.85\linewidth}{\centering
    Junze Liu$\mathrm{^{\ast1}}$, Aishik Ghosh$\mathrm{^{\ast1,2}}$, Dylan Smith$\mathrm{^{\ast1}}$, Pierre Baldi$\mathrm{^1}$, Daniel Whiteson$\mathrm{^1}$}\vspace{2mm}\\
  $^\mathrm{1}$University of California, Irvine\\
  $^\mathrm{2}$Lawrence Berkeley National Laboratory, Berkeley\\
  \parbox[c]{0.88\linewidth}{\centering 
    \texttt{\{junzel1, aishikg, dylanrs, pfbaldi, daniel\}@uci.edu}} \\
}

\begin{document}

\maketitle
\def\thefootnote{*}\footnotetext{These authors contributed equally to this work}

\begin{abstract}
  Calorimeter shower simulations are often the bottleneck in simulation time for particle physics detectors. A lot of effort is currently spent on optimizing generative architectures for specific detector geometries, which generalize poorly. We develop a geometry-aware autoregressive model on a range of calorimeter geometries such that the model learns to adapt its energy deposition depending on the size and position of the cells. This is a key proof-of-concept step towards building a model that can generalize to new unseen calorimeter geometries with little to no additional training. Such a model can replace the hundreds of generative models used for calorimeter simulation in a Large Hadron Collider experiment. For the study of future detectors, such a model will dramatically reduce the large upfront investment usually needed to generate simulations.
\end{abstract}
\nsection{Introduction}
Deep generative models have become essential in solving the fast simulation needs in particle physics~\cite{ATL-SOFT-PUB-2018-001,Butter:2020tvl,Deja:2019vcv,Chekalina:2018hxi,ATLAS:2021pzo}. Of particular concern are calorimeter simulations, which take up a large component of the computing budget of physics experiments~\cite{ATLAS:2021pzo}. While the same base architecture is found to perform reasonably on a multitude of natural image datasets, this has not been the case for calorimeter data. Calorimeter data are three dimensional images where the resolution of the image is determined by the size of the calorimeter cells (analogous to pixels). These cells have varying shapes and sizes even within the same calorimeter, as well as between detectors from different experiments. A significant amount of research time and computing resources are spent to develop generative architectures from scratch for different calorimeter geometries~\cite{Buhmann:2020pmy,Krause:2021ilc,Mikuni:2022xry,Khattak:2021ndw}. These studies have usually ignored challenging transition regions of the calorimeter where the geometry abruptly changes (for example the cells become wider in one dimension at a transition point), or resorted to training hundreds of networks to simulate different geometries within the same calorimeter~\cite{ATLAS:2021pzo}. Developing an architecture that can generalize to multiple calorimeters would significantly cut down on the R\&D time  invested by various experiments to develop geometry specific models from scratch. Such a generalizable architecture would make future detector studies cheaper by dramatically reducing the upfront computational investment required to simulate a new detector design. These studies currently rely on slow first principles based simulation software (Geant~\cite{GEANT4:2002zbu,Allison:2016lfl,Allison:2006ve}), but a general purpose foundational (pre-trained) generative model could reduce the demand for such simulations. When the difference in geometry is only in the segmentation of the detector, the model could adapt to the new geometry without any additional training (zero-shot adaptation), and for calorimeter designs with a different composition of interaction material, it could act as a foundational model that is fine-tuned to the new calorimeter with far fewer training samples compared to what would be required to develop a model from scratch.

To solve this task, we investigate a geometry-aware autoregressive generative model which learns to adapt its energy deposition from cell to cell depending on its size and position, as well as the energy distribution in its neighbouring cells. Deep autoregressive models (ARMs) have demonstrated impressive performance for modeling the sparsity in images, thus are well suited for this task~\cite{van2016pixel,salimans2017pixelcnn++,lu2021sparse}. The model is trained on a range of calorimeter geometries, including the particularly challenging transition regions. We show that our model can adapt to these differences and reproduce the impact the geometry of a calorimeter has on important physics variables. This is a key step towards building a generalizable architecture that can adapt to new unseen calorimeter designs with little to no additional training. 
\nsection{Dataset}

The training dataset is generated using Geant4~\cite{GEANT4:2002zbu,Allison:2016lfl,Allison:2006ve} with the calorimeter design created originally for CaloGAN dataset~\cite{Paganini_2018}, but we modified the segmentation of the detector for our study. The calorimeter has three layers, `inner', `middle', and `outer', denoted by $z= \{1,2,3\}$ and two orthogonal coordinates, $\eta$ and $\phi$. It records a three dimensional image but with no time information. To generate particle showers, we shoot single photons at the centre ($\eta=0$, $\phi=0$) of the calorimeter. The photon generates a shower of particles as it interacts with the material, and the shower develops sequentially through the calorimeter layers. 10,000 particle showers for photons with 65~GeV of energy were generated at (pseudo-)point-cloud level using Geant4 and then binned into cells of varying granularities in $\eta$-$\phi$ space. The innermost layer is generated by starting from the default dimension of (12, 12) and uniformly sampling multiplicative factors along each dimension: 4 and 16 in $\eta$ and $\sfrac{1}{3}$, 1, 2, and 4 in $\phi$. The middle calorimeter layer was generated by uniformly sampling layer sizes of (12, 12), (48, 24), (48,48), and (36, 48), wherein the final geometry has 12 cells on the left half of the image and 24 cells on the right half in $\eta$ (this represents the transition region of a full calorimeter) ; all other geometries have uniform cell sizes across each image. Finally, the outermost layer are all of the same shape, (24, 24), as this is the sparsest layer. These geometries were chosen to resemble the range of calorimeter segmentation that exist in the ATLAS experiment~\cite{ATLAS:2008xda} electromagnetic calorimeter.
\nsection{Methods}

\label{methods}
\begin{figure}
  \centering
  \includegraphics[width=.5\textwidth]{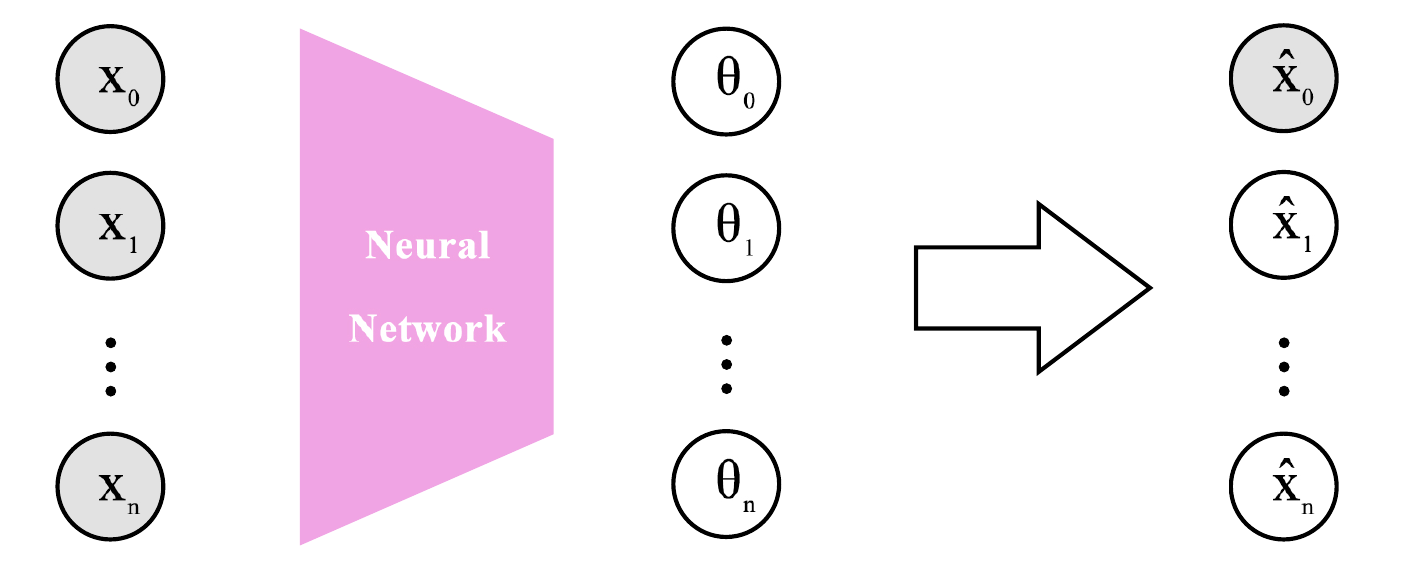}
  \ncaption{The architecture of one out of three Discrete Mixture Model Autoregressive Model (D+D ARM) we implement. The true starting value $\mathrm{x_0}$ is given when generating the samples. $\mathrm{\theta_{i}}$ is learned during the training and used to sample the energy deposit of each cell in calorimeter simulation.}
  \label{architecture}
\end{figure}

\begin{figure}
  \centering
  \subfloat[D+D ARM with cell sizes as extra inputs]{\includegraphics[width=.44\textwidth]{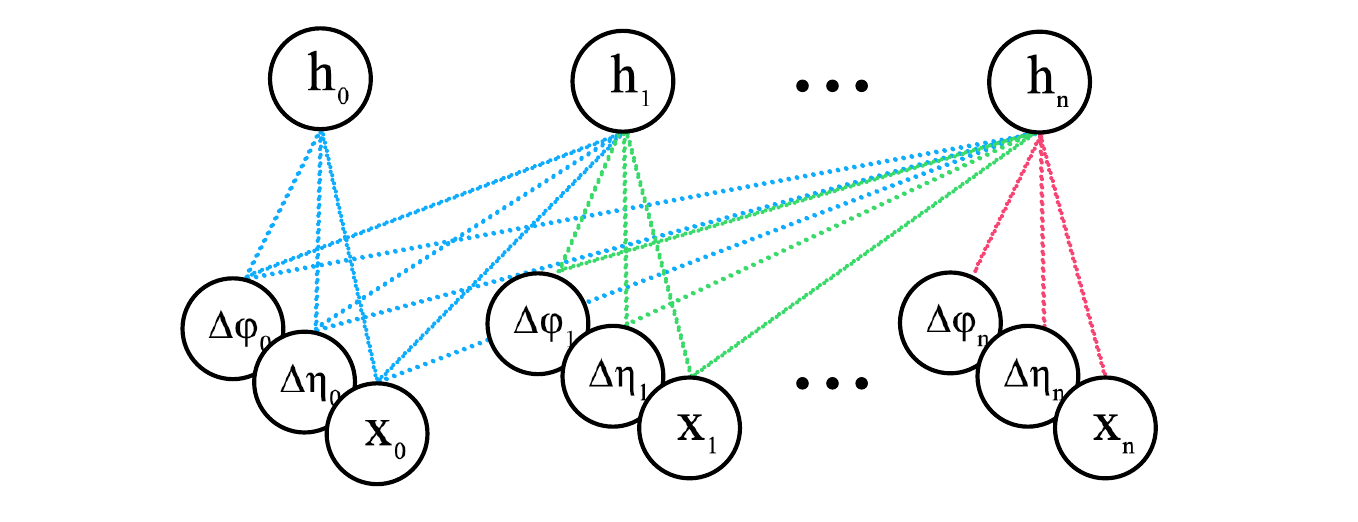}\label{withcellsize}}\hspace{.1\textwidth}
  \subfloat[The spiral path]{\includegraphics[width=.22\textwidth]{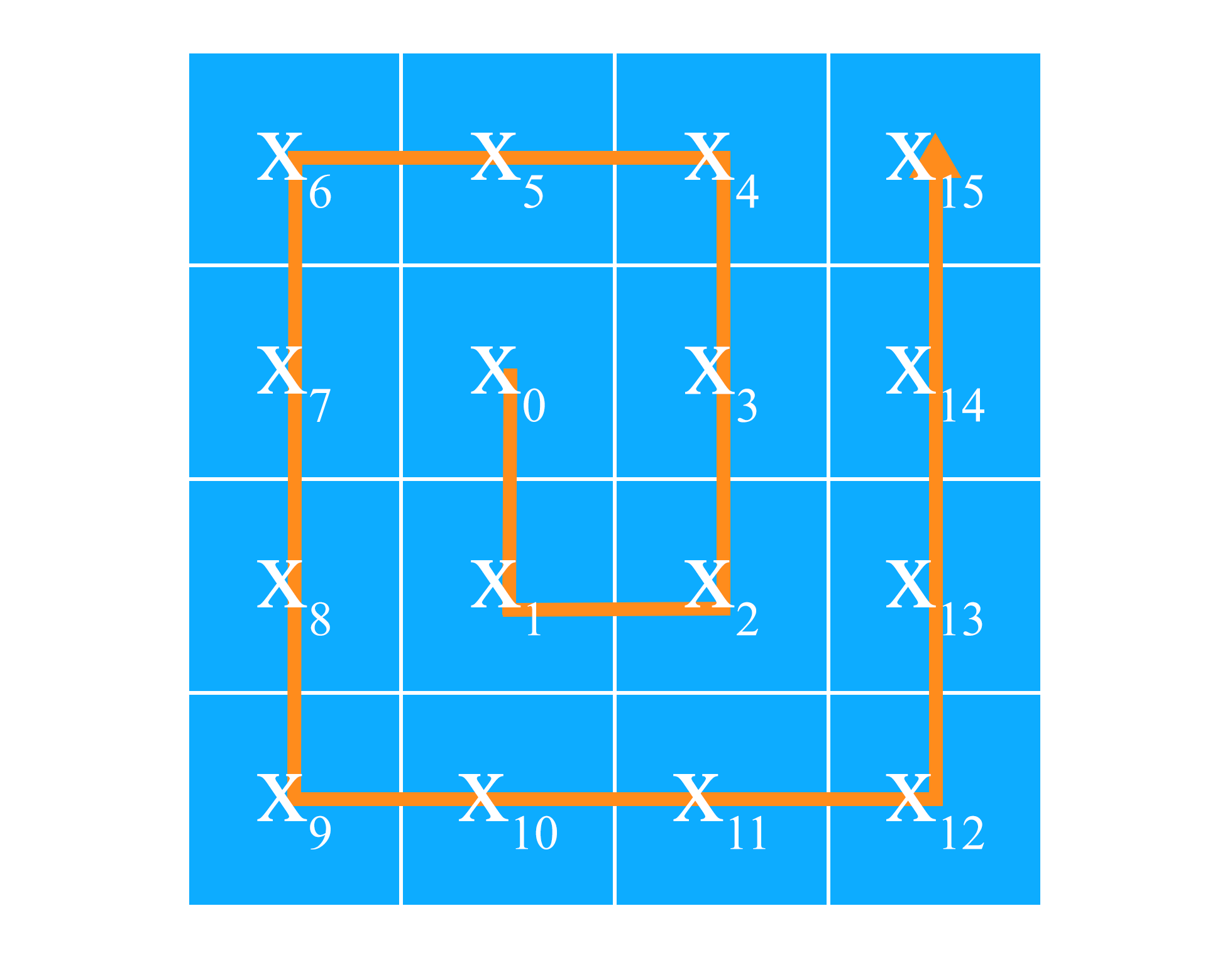}\label{spiralpath}}
  \ncaption{(a) The cell's size in $\eta$ and $\phi$ is added as the extra input to provide the geometry information. All three inputs of each cell (energy deposit, size in $\eta$, and size in $\phi$) are aggregated to hidden nodes according to the input ordering; (b) the 2D energy deposit matrix is flattened using a spiral path.}
  \label{data_prep}
\end{figure}

We design a framework based on the autoregressive model (ARM) as an efficient data generator for simulated multi-layer calorimeter data. The framework is composed of three Discrete (D+D) ARMs trained separately for each layer (Figure~\ref{architecture}). Each D+D ARM is built based on Masked Autoencoder for Distribution Estimation (MADE)~\cite{germain2015made}. While traditional autoregressive models generate each cell in the output sequentially, MADE generates all desired parameters with a single pass through the regular autoencoder. This feature allows us to exploit the parallel computation on GPUs for faster training. 
The D+D ARM for the generation of the inner layer consists of one masked fully connected layer and one 1D convolutional layer, and the D+D ARMs for the generation of the middle layer and the outer layer consist of five masked fully connected layers and one 1D convolutional layer. We use GELU~\cite{hendrycks2016gaussian} as the activation function. Then, we have a softmax layer to classify the value into N categories, where N is the closest greater integer of the max cell value in the sample. Each generated energy deposit value is modeled as a categorical distribution.

\begin{figure}[t!]
  \centering
  \subfloat[Inner layer - (192, 12)]{\includegraphics[width=.40\textwidth]{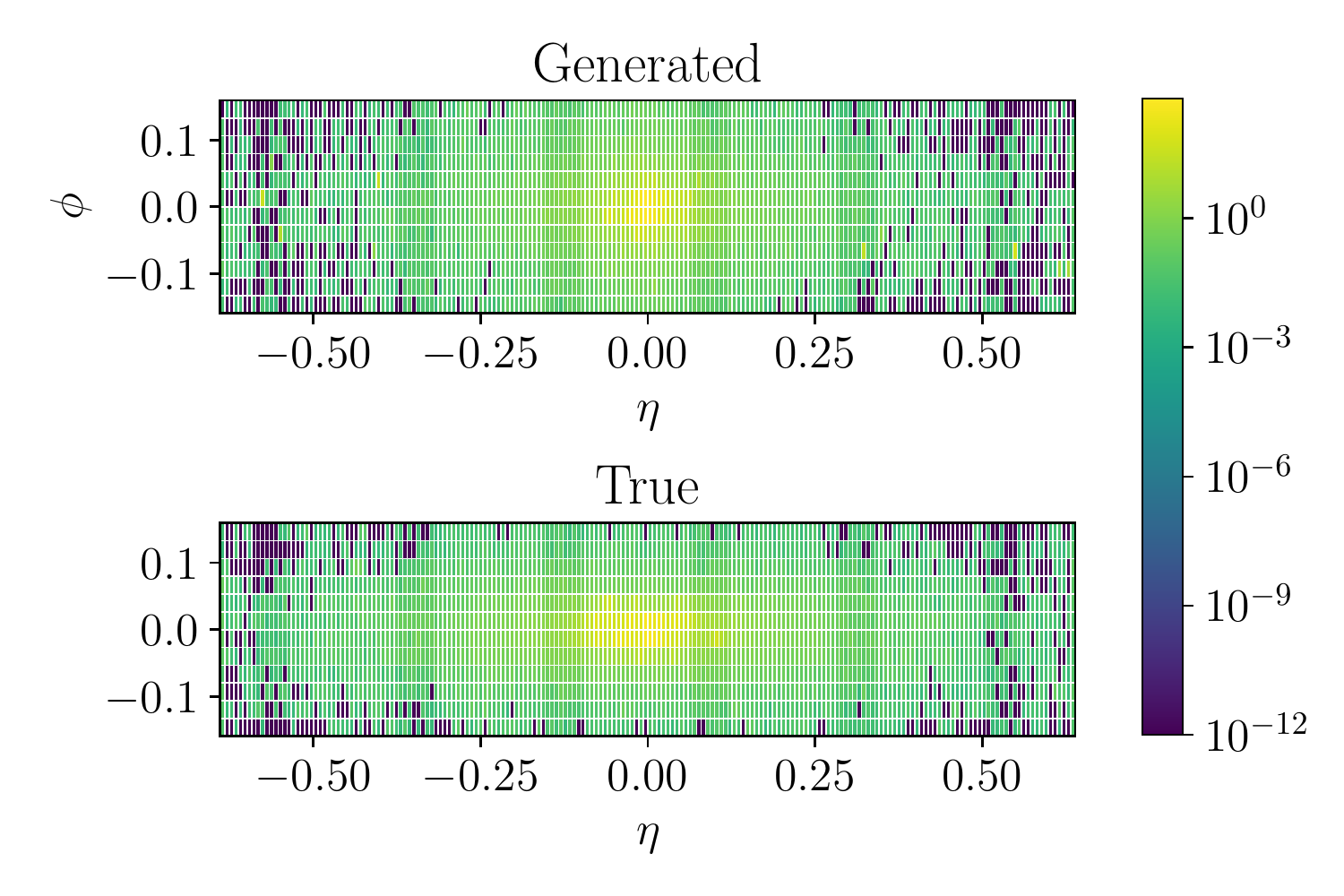}\label{meanimage_stripelayer}}\hfill
  \subfloat[Middle layer - (36, 48)]{\includegraphics[width=.5\textwidth]{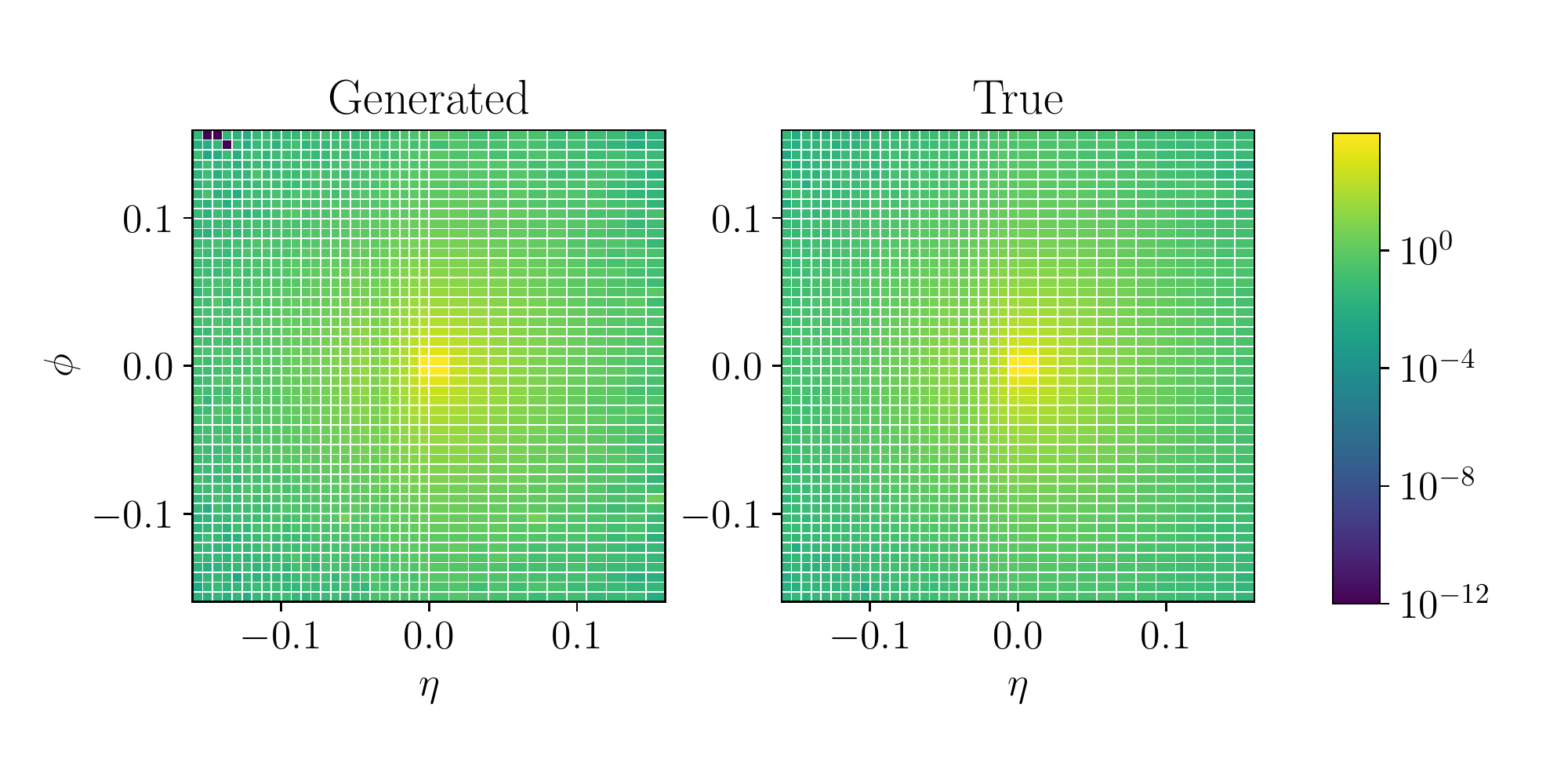}\label{meanimage_middlelayer}}\hfill
  \ncaption{The average generated data is similar to the average original data. (a) The generated and true average inner layer of (192, 12), and (b) the middle layer of (36, 48).}
  \label{average_image}
\end{figure}

In addition to the original energy deposits, the cell sizes in $\eta$ and $\phi$ dimensions are also given as inputs to the D+D ARM, enabling the model to adaptively generate the energy distribution for different geometries (Figure~\ref{withcellsize}). During training, the inputs of each ARM includes a matrix of the original simulated calorimeter data with energy deposits in cells and two matrices of each cell's size in $\eta$ and $\phi$. We start the generation from the central cell because the shower typically deposits the most energy near the center, and this acts as a good starting point for the ARMs. In addition, we give the cell size ($\Delta\eta,\Delta\phi$) as additional inputs, where $\Delta\eta$ is the length along the $\eta$-axis and $\Delta\phi$ is the length along the $\phi$-axis. This allows the ARMs to learn the energy distribution as a function of these geometry features. At test time, we give the networks a value for the central cell as a starting point, which is sampled from a known prior distribution, and two matrices which describe the size of each cell in $\eta$ and $\phi$. The trained D+D ARM then generates the energy deposits in the rest cells by sampling from the learnt multinomial probability distribution.

The calorimeter data is carefully prepared to fit in our models. The 2D matrix is flattened using a spiral path counterclockwise to fit in our D+D ARM, so the neighboring cells share similar energy deposit values (Figure~\ref{spiralpath}). We discretize the energy deposit value by rounding it to the nearest integer. The energy depositions vary by several orders of magnitude from one cell to another, and preprocessing them by a power transformation, $\hat{x}=x^{1/p}$, was found to improve training. The optimal value for the hyper-parameter $p$ was found to be $p=2$. We split the dataset into training, validation, and test sets in the ratio 0.8:0.1:0.1. The models were trained in PyTorch~\cite{NEURIPS2019_9015} up to 100 epochs using the Adam optimizer~\cite{kingma2014adam} and a negative log likelihood as the loss function on two NVIDIA RTX 3090 GPUs. The performance is evaluated on the test set by computing physics observables from the raw Geant4 and ARM generated images and comparing their distributions. The physics variables include energy weighted mean and standard deviations in $\eta$ and $\phi$ directions, and the correlation between lateral ($\eta$, $\phi$) distributions between the calorimeter layers in the $z$ direction. We also visually inspect individual samples, and the average images for each geometry.

\nsection{Results}

% \begin{figure}[h!]
%   \centering
%   \subfloat[Inner layer - (48, 24)]{\includegraphics[width=.5\textwidth]{figures/energy-weighted_avgs/energy-weighted_avgs_stripe_layer_(48, 24).pdf}\label{eng-weighted_avgs_stripelayer}}\hfill
%   \subfloat[Inner layer - (192, 12)]{\includegraphics[width=.5\textwidth]{figures/energy-weighted_avgs/energy-weighted_avgs_stripe_layer_(192, 12).pdf}\label{eng-weighted_avgs_stripelayer}}\hfill\\
%   \subfloat[Middle layer - (36, 48)]{\includegraphics[width=.5\textwidth]{figures/energy-weighted_avgs/energy-weighted_avgs_middle_layer_(36, 48).pdf}\label{eng-weighted_avgs_middlelayer}}\hfill
%   \subfloat[Middle layer - (48, 48)]{\includegraphics[width=.5\textwidth]{figures/energy-weighted_avgs/energy-weighted_avgs_middle_layer_(48, 48).pdf}\label{eng-weighted_avgs_middlelayer}}\hfill
%   \caption{Distribution of the energy weighted means of (a) inner layer, (b) middle layer, and (c) outer layer. The generated data and the original data share a similar distribution.}
%   \label{energy-weighted_avgs}
% \end{figure}
\begin{figure}[t!]
  \centering
  \subfloat[Middle layer - (36, 48)]{\includegraphics[width=.5\textwidth]{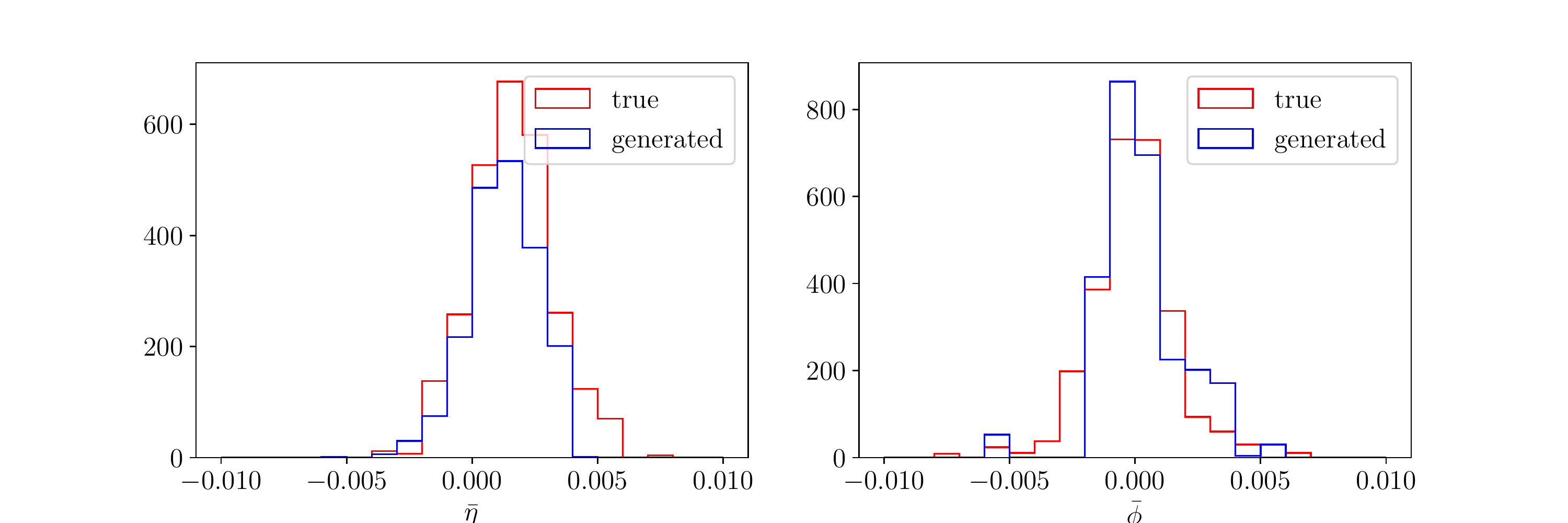}\label{eng-weighted_avgs_middlelayer}}\hfill
  \subfloat[Middle layer - (48, 48)]{\includegraphics[width=.5\textwidth]{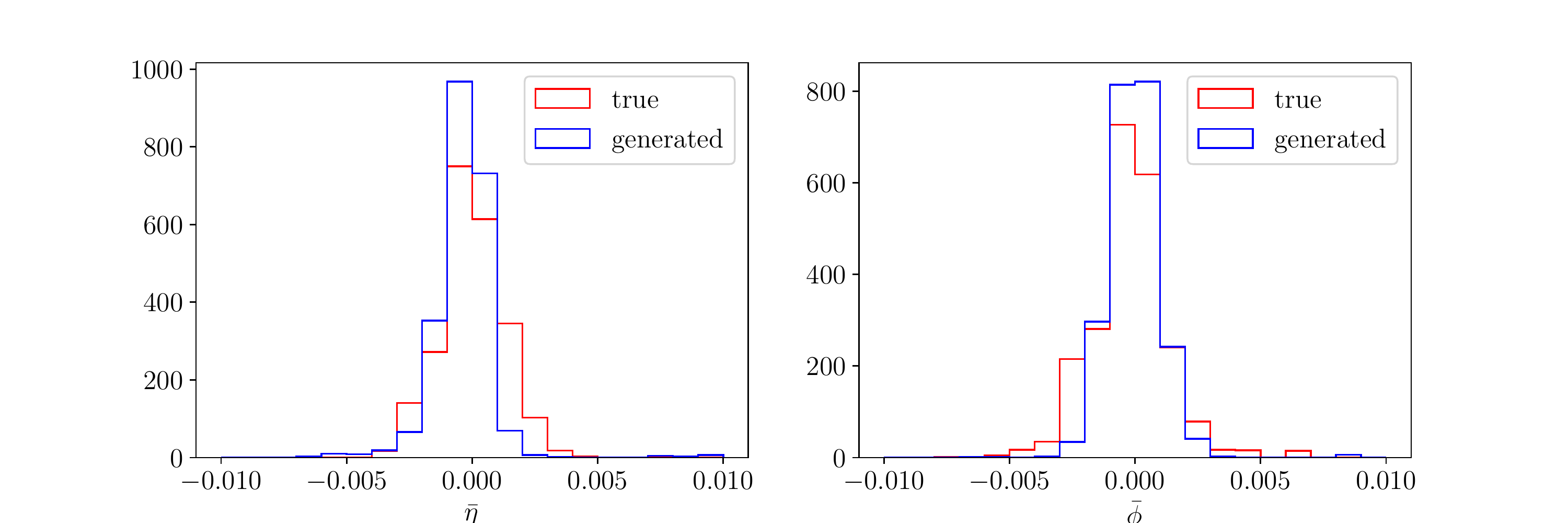}\label{eng-weighted_avgs_middlelayer}}\hfill
  \ncaption{Distribution of the energy weighted means of samples of (a) shape (36, 48) and (b) shape (48, 48) in the middle layer. The generated data and the original data share a similar distribution.}
  \label{energy-weighted_avgs}
\end{figure}

% \begin{figure}[h!]
%   \centering
%   \subfloat[Inner layer - (48, 24)]{\includegraphics[width=.5\textwidth]{figures/shower_widths/shower_widths_stripe_layer_(48, 24).pdf}\label{shower_widths_stripelayer}}\hfill
%   \subfloat[Inner layer - (192, 12)]{\includegraphics[width=.5\textwidth]{figures/shower_widths/shower_widths_stripe_layer_(192, 12).pdf}\label{shower_widths_stripelayer}}\hfill\\
%   \subfloat[Middle layer - (36, 48)]{\includegraphics[width=.5\textwidth]{figures/shower_widths/shower_widths_middle_layer_(36, 48).pdf}\label{shower_widths_middlelayer}}\hfill
%   \subfloat[Middle layer - (48, 48)]{\includegraphics[width=.5\textwidth]{figures/shower_widths/shower_widths_middle_layer_(48, 48).pdf}\label{shower_widths_middlelayer}}\hfill
%   \caption{Distribution of the shower widths of (a) inner layer, (b) middle layer, and (c) outer layer. The generated data and the original data share a similar distribution.}
%   \label{shower_widths}
% \end{figure}
% \begin{figure}[h!]
%   \centering
%   \subfloat[Middle layer - (36, 48)]{\includegraphics[width=.5\textwidth]{figures/shower_widths/shower_widths_middle_layer_(36, 48).pdf}\label{shower_widths_middlelayer}}\hfill
%   \subfloat[Middle layer - (48, 48)]{\includegraphics[width=.5\textwidth]{figures/shower_widths/shower_widths_middle_layer_(48, 48).pdf}\label{shower_widths_middlelayer}}\hfill
%   \caption{Distribution of the shower widths of (a) samples of shape (36, 48) and (b) samples of shape (48, 48) in the middle layer. The generated data and the original data share a similar distribution.}
%   \label{shower_widths}
% \end{figure}

\begin{figure}[t!]
  \centering
  \subfloat[Inner (48, 12) \& middle (12, 12) layer]{\includegraphics[width=.5\textwidth]{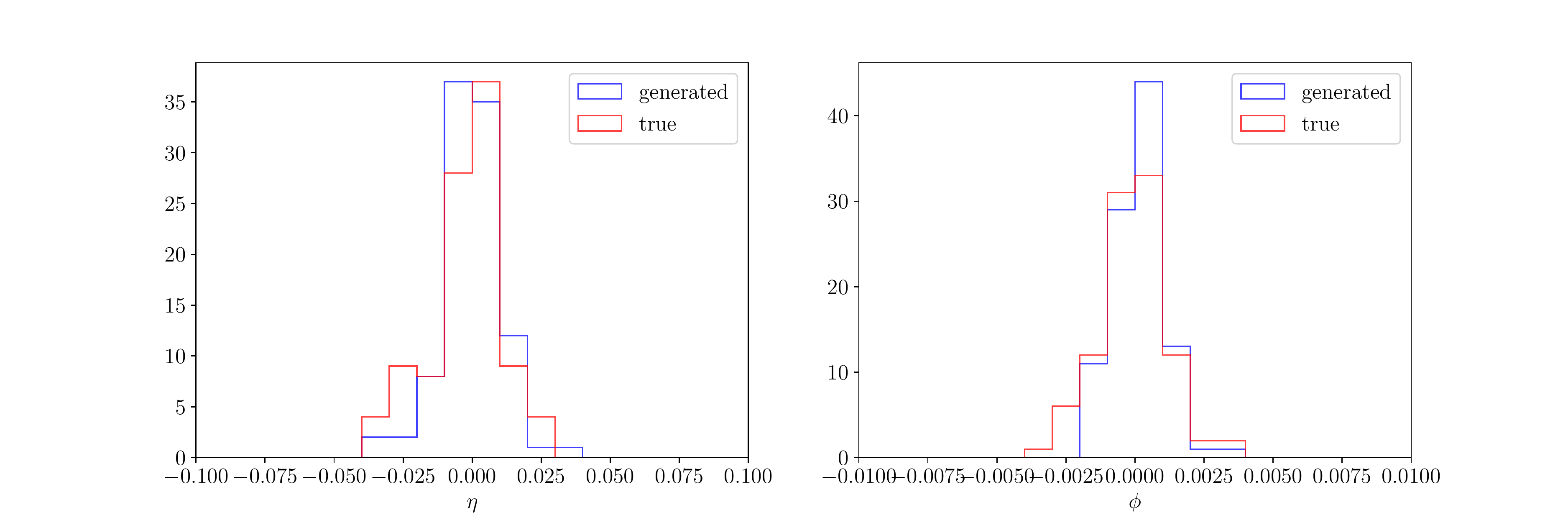}\label{shower_widths_middlelayer}}\hfill
  \subfloat[Inner (48, 4) \& middle (36, 48) layer]{\includegraphics[width=.5\textwidth]{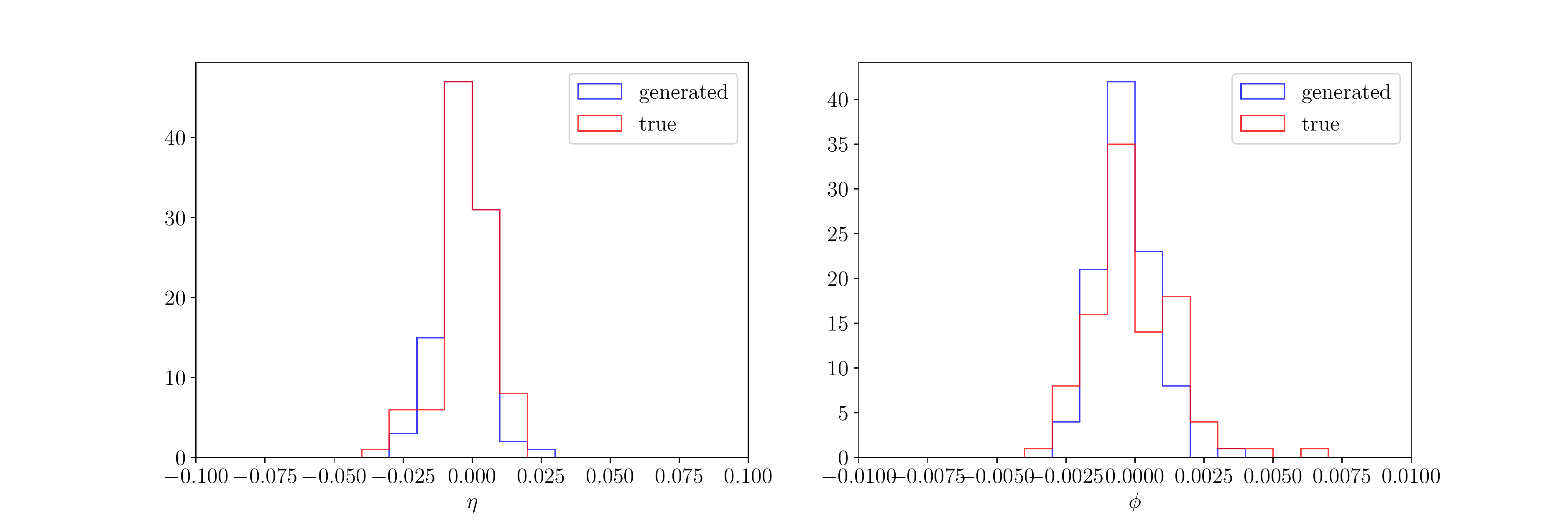}\label{shower_widths_middlelayer}}\hfill
  \ncaption{Distribution of the distance of energy weighted means between (a) samples of (48, 12) in the inner layer and (12, 12) in the middle layer, and (b) samples of shape (48, 4) in the inner layer and (36, 48) in the middle layer. The generated data and the original data share a similar distribution.}
  \label{layer_correlation}
\end{figure}

Our D+D ARMs show promising performance in the ability to adapt data generation to various calorimeter geometries. We study the performance in two aspects: qualitative and quantitative analyses. For qualitative analysis, we visualize the average of all generated samples for each calorimeter geometry. The average sample of each layer generated by our D+D ARMs is compared with the original average sample. Figure~\ref{average_image} shows two example geometries, (192, 12) and (36, 48), from the inner layer and middle layer respectively. The samples of (36, 48) from the middle layer are particularly challenging because they have nonuniform cell sizes along the $\eta$ direction. The results show that our geometry-aware D+D ARMs are able to learn to generate data with more energy deposit in larger cells while less in smaller cells within the same sample. For quantitative analysis, we show three physics variables: energy weighted means, shower widths, and correlations between layers. The energy weighted mean is summed over all cells in an image as Eq.~\ref{eq_energy_weighted_mean} in $\eta$ (and likewise in $\phi$):
\begin{equation}\footnotesize
\label{eq_energy_weighted_mean}
\bar{\eta} = \frac{\sum_{i} \eta_{i}E_{i}}{\sum_{i} E_{i}}
\end{equation}
where $E_{i}$ is the energy deposit
% and $\eta_{i}$ is the cell location in $\eta$
of the $i^{th}$ cell. The shower width is as Eq.~\ref{eq_shower_width} in $\eta$ (and likewise in $\phi$):
\begin{equation}\footnotesize
\label{eq_shower_width}
\sigma_{\eta} = \sqrt{\frac{\sum_i E_{i} (\eta_i-\bar{\eta})^2}{\frac{(M-1)}{M} \sum_{i} E_{i}}}
\end{equation}
where $M$ is the number of cells with non-zero energy. The distributions of these two parameters calculated from all generated and original samples of each geometry are visualized side by side to study the performance of our D+D ARMs in statistics (Figure~\ref{energy-weighted_avgs}). The slight off-set to the right of the peak in the $\eta$ distribution is a feature of transition regions of calorimeters and therefore expected for the geometry (36, 48) of the middle layer. 
%The central cell in the right half has $\eta$=0.0125 but the central cell on the left is only -0.00625 which pushes the avg slightly to the right. 
We choose samples of the shape (36, 48) and (48, 48) from the middle layer to show the robustness of our D+D ARMs in the generation of different geometries regardless of the varying cell size. At last, we show the correlation of samples between two layers with different geometries by calculating the distance between their energy weighted means (Figure \ref{layer_correlation}). The distributions of correlations are visualized between samples of (48, 12) from the inner layer and (12, 12) from the middle layer, as well as samples of (48, 4) from the inner layer and (36, 48) from the middle layer. The generated samples share similar distribution as the original samples, indicating that our D+D ARMs learns the correlation between layers.

% \begin{table}
%   \caption{Sample table title}
%   \label{sample-table}
%   \centering
%   \begin{tabular}{lll}
%     \toprule
%     \multicolumn{2}{c}{Part}                   \\
%     \cmidrule(r){1-2}
%     Name     & Description     & Size ($\mu$m) \\
%     \midrule
%     Dendrite & Input terminal  & $\sim$100     \\
%     Axon     & Output terminal & $\sim$10      \\
%     Soma     & Cell body       & up to $10^6$  \\
%     \bottomrule
%   \end{tabular}
% \end{table}
\nsection{Conclusion and Future Directions}
\label{conclusion}

We study the first attempt of training a geometry-aware generative model to simulate a range of calorimeter geometries and demonstrate that the model is able to adapt its energy in response to different calorimeter geometries. This is the first step towards building a general purpose generative architecture that can handle new calorimeter segmentations and become a pre-trained base that can be quickly tuned to new calorimeter designs.  Our proposed geometry-aware autoregressive model shows promising performance for this challenging task, with room for further improvement. The model learns a generalized way to simulate different calorimeter geometries by learning to use information about each individual cell; the cell size and position. Our experiments prove that deep autoregressive models are able to adaptively learn to generate the energy distribution when the cell size varies even within individual calorimeter layers. The method we implemented is still preliminary but shows us the direction for the next phase. Further work will focus on improving the fidelity of the generated images. We may enforce sparsity through Sparse Autoregressive Models (SARMS)~\cite{lu2021sparse} that use an extra parameter to model the sparsity in the data. We may also try to generate continuous outputs, instead of discrete outputs we currently have. While we currently sample the energy of the central cell from a histogram, in the future this may be generated with a neural network. By making the generation of each layer fully conditional on other layers, we could further improve the correlation between the generated layers. Finally, we will study the zero-shot adaptation of the model for more and more challenging geometries.
\section*{Acknowledgement}
\label{sec:ack}
JZ and PB are in part supported by grant ARO  76649-CS to PB. AG, DS, and DW are supported by the U.S. Department of Energy, Office of Science under grant DE-SC0009920.  DS is additionally supported by a HEPCAT fellowship from the U.S. Department of Energy.
\printbibliography

\clearpage

\end{document}